\documentclass[]{fairmeta_gr2}

\usepackage{amsmath}
\usepackage{amssymb}
\usepackage{pifont}
\usepackage[flushleft]{threeparttable}
\usepackage{tabularx}
\usepackage{algorithm}
\usepackage{algorithmic}
\usepackage{xspace}
\usepackage{array}

\title{The Efficiency vs. Accuracy Trade-off: Optimizing RAG-Enhanced LLM Recommender Systems Using Multi-Head Early Exit}

\author[1,2]{Huixue Zhou}
\author[3]{Hengrui Gu}
\author[2]{Zaifu Zhan}
\author[1]{Xi Liu}
\author[3]{Kaixiong Zhou}
\author[1]{Mingfu Liang}
\author[2]{Yongkang Xiao}
\author[1]{Srinivas Govindan}
\author[1]{Piyush Chawla}
\author[1]{Jiyan Yang}
\author[1]{Xiangfei Meng}
\author[1]{Huayu Li}
\author[1]{Buyun Zhang}
\author[1]{Liang Luo}
\author[1]{Wen-Yen Chen}
\author[1]{Yiping Han}
\author[1]{Bo Long}
\author[2]{Rui Zhang}
\author[4]{Tianlong Chen}

\affiliation[1]{Meta Platforms}
\affiliation[2]{University of Minnesota}
\affiliation[3]{NCSU}
\affiliation[4]{UNC at Chapel Hill}

\abstract{The deployment of Large Language Models (LLMs) in recommender systems for Click-Through Rate (CTR) prediction requires a careful balance between computational efficiency and predictive accuracy. This paper introduces \textbf{OptiRAG-Rec}, a comprehensive framework that integrates Retrieval-Augmented Generation (RAG) with a novel multi-head early exit architecture to address both challenges. By leveraging Graph Convolutional Networks (GCNs) as efficient retrieval mechanisms, the framework significantly reduces data retrieval times while maintaining high model performance. Additionally, the multi-head early exit strategy dynamically terminates inference based on real-time predictive confidence assessments, enhancing responsiveness without sacrificing accuracy. Experimental results demonstrate that \textbf{OptiRAG-Rec} reduces computation time while preserving the precision required for reliable recommendations, establishing a new benchmark for efficient and accurate LLM deployment in recommendation.
\footnote{Disclaimer: No internal or proprietary Meta data was used in this study.}}

\date{June 1, 2025}
\correspondence{Huixue Zhou at \email{zhou1742@umn.edu}; Xi Liu at \email{xliu1@meta.com}; Tianlong Chen at \email{tianlong@cs.unc.edu}}

\begin{document}

\maketitle

\section{Introduction}

\begin{figure}[t]
  \includegraphics[width=.9\linewidth]{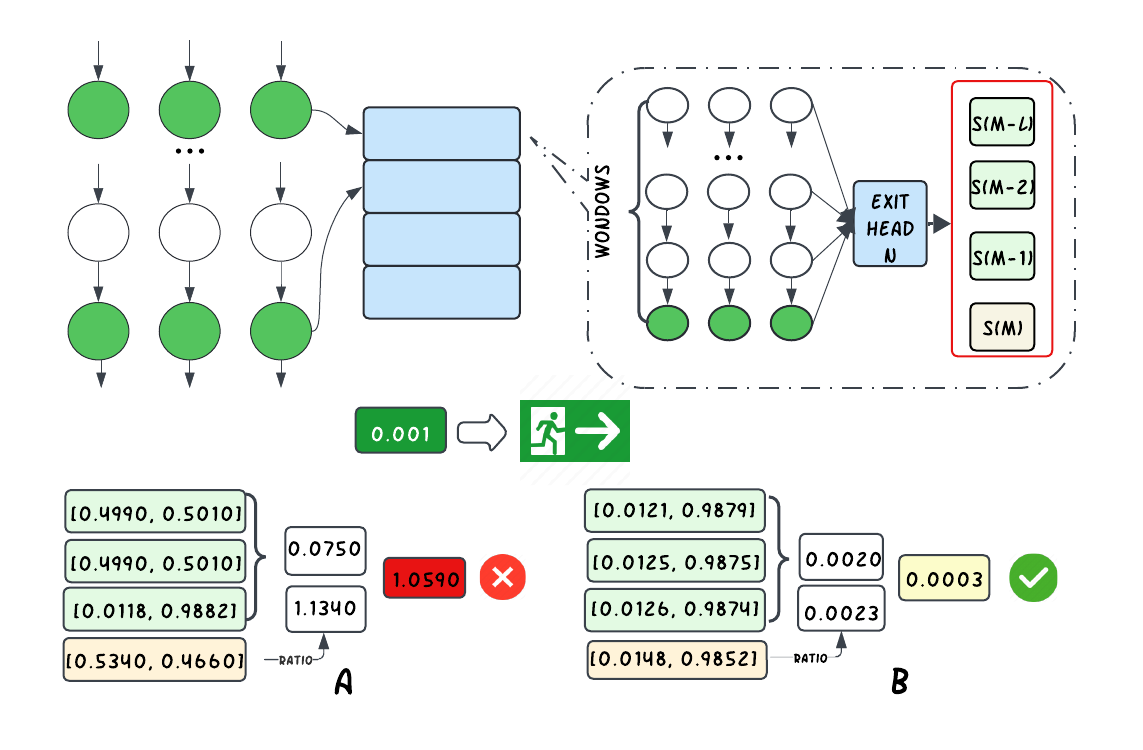}
  \caption{An example of multi-head early exit based on ratio scores of positive and negative predictions: \textbf{A}: No early exit due to the significant difference between the average ratio of previous layers and the exit layer exceeding the threshold (0.001). \textbf{B}: Early exit achieved with minimal changes in ratio scores.}
  \label{fig:early_exit_example}
\end{figure}
Due to their remarkable capabilities in semantic understanding and knowledge retention, Large Language Models (LLMs) have demonstrated impressive performance across various domains, becoming essential components of various text-based recommendation systems, such as sequential recommendation~\cite{bert4rec_wang,sequence_harte2023leveraging,sequence_li2023prompt,sequence_zheng2024harnessing} and ranking~\cite{rank_zhao2024let,rank_acharya2023llm}. In light of this, numerous researchers have sought to adapt LLMs for Click-Through Rate (CTR) prediction problem~\cite{WangLim2023,BaoEtAl2023,LinEtAl2024}, leveraging their text-mining capabilities to analyze textual user behaviors for more accurate preference modeling.

The performance of LLM-based CTR predictors is enhanced by Retrieval-Augmented Generation (RAG) modules \cite{Rella_Lin_2024,hajiaghayi2024adauctionsllmsretrieval}, which integrate diverse and user-relevant texts into the input contexts. This enrichment allows LLMs to extract more accurate user features for preference prediction.  

However, the integration of RAG modules into LLMs introduces significant efficiency challenges, particularly concerning the Inference efficiency. 
Two critical bottlenecks arise from the introduction of RAG modules: 
\ding{182} \textbf{Retrieval Efficiency}: The RAG framework adds a multi-stage process involving encoding, retrieval, and prefilling steps for retrieved contexts. This sequential execution causes substantial delays due to the time required for retrieval, thus hindering the prompt commencement of the inference process.
\ding{183} \textbf{Inference Overhead}: With extended input lengths, the computational demand surges, exacerbating the model's inference time due to the quadratic time complexity of LLMs relative to input length. This significant increase in computational overhead severely impacts the real-time responsiveness of the model, limiting its practical applicability in real-world scenarios.

In the realm of enhancing LLM inference efficiency, prevalent techniques such as model compression through quantization and Fast Decoding Algorithms like Speculative Decoding (e.g., Medusa\cite{cai2024medusa} and Kangaroo\cite{liu2024kangaroo}) are widely employed. These methods aim to reduce model size and accelerate the decoding process, thereby shortening overall inference times. Despite their benefits, these strategies often come with trade-offs, particularly in terms of model accuracy\cite{speculative_xia2024unlocking,cai2024medusa,jin2024comprehensive}. Quantization, for instance, achieves faster processing by reducing the precision of computations, which can lead to the loss of critical data details. Similarly, Fast Decoding Algorithms may prioritize speed over optimality, potentially selecting suboptimal predictive paths in their rush to generate quick responses. While these approaches address certain aspects of inference efficiency, they do not fully resolve the bottlenecks introduced by RAG modules, such as retrieval delays and increased computational overhead from extended input lengths. Consequently, there remains a need for more holistic solutions that balance efficiency, accuracy in RAG-enhanced LLMs.

To address these challenges, we propose \textbf{OptiRAG-Rec}, a novel framework designed to enhance the efficiency of LLM-based CTR while maintaining high recommendation quality. This framework integrates two innovative acceleration techniques tailored to the identified bottlenecks: retrieval efficiency and inference overhead.

First, to address retrieval efficiency (Bottleneck \ding{182}), we introduce GCN-Retriever, a lightweight yet highly effective retrieval scheme. Leveraging the capabilities of Graph Convolutional Networks (GCNs)~\citep{gcn1,gcn2}, GCN-Retriever models structural data in user-item graphs, capturing multi-order interaction information to generate precise and informative representations. By replacing computationally expensive LLM embedders, GCN-Retriever significantly reduces retrieval times, minimizing inference delays while maintaining high recommendation performance. Second, to tackle inference overhead (Bottleneck \ding{183}), we propose a multi-head early exit strategy integrated with an exit scoring mechanism tailored for CTR prediction tasks. This strategy dynamically terminates the inference process when predictions meet a confidence threshold, reducing unnecessary computational overhead. By doing so, the system accelerates response times without sacrificing recommendation accuracy.
The components of \textbf{OptiRAG-Rec}—GCN-Retriever, the LLM-based recommender, and the multi-head early exit mechanism—are tightly integrated to address specific challenges: GCN-Retriever enhances retrieval efficiency using graph-based representations, the LLM-based recommender ensures precise predictions, and the multi-head early exit mechanism reduces computational overhead during inference. Together, they synergistically improve the recommender system's performance by balancing efficient data retrieval, accurate predictions, and optimized throughput. Below, we summarize the key contributions of our framework:
\begin{itemize}
    \item \textbf{Enhanced LLM for CTR Prediction via RAG}: Improves CTR accuracy by integrating interaction data into  LLM models.
    \item \textbf{Efficient GCN-Retriever}: Introduces a lightweight GCN-based retrieval mechanism, reducing retrieval latency without compromising recommendation quality.
    \item \textbf{Inference Time Optimization through Early Exit}: Implements a dynamic early exit strategy, reducing computational costs by terminating inference at intermediate layers when predictions meet confidence thresholds.
    \item \textbf{Novel Multi-Head Early Exit Adjustment}: Proposes a multi-head architecture for early exiting, maintaining accuracy while significantly improving inference efficiency.
\end{itemize}

\section{Related Work}
\subsection{Language Models for Recommendation}
Building on prior research\cite{llm4rec_wu2023survey}, the integration of language models into recommender systems often focuses on their distinct functions within the recommendation process. These roles include serving as feature extractors\cite{llm4rec_feature_Bian2022,llm4rec_feature_zheng2023generative,llm4rec_feature_zhang2024generativeagentsrecommendation}, where language models analyze item and user data to produce embeddings or tokens. These embeddings can be utilized by traditional recommender system models to enhance task-specific recommendations through knowledge-aware embeddings.

Furthermore, language models can function within scoring or ranking mechanisms\cite{bert4rec_wang,llm4rec_Zhu_2024,llm4ec_asranker_kim2024}. This approach leverages pre-trained language models to transform recommendation systems significantly. Typically, the input sequence includes task instructions, behavioral prompts, with the output generating pertinent recommendation results.

Our methodology diverges from previous practices by employing the language model primarily within the scoring functions, while using a simple traditional model, specifically a GNN, as a retriever. This model extracts similar user profiles to construct the prompts for the language model, capitalizing on its ability to comprehend and synthesize user data and interactions, thereby generating personalized recommendations.

\subsection{Efficient Inference}
The inference performance of LLMs is often constrained by the sequential nature of auto-regressive decoding, where the generation of each token necessitates a full network forward pass. To address the high inference latency inherent in LLMs, several strategies have been proposed: Techniques such as quantization \cite{quan_Fan2020,quan_Bai2022,quan_Tao2022}, pruning \cite{puring_ma2023llm,puring_sun2023simple,puring_xia2023sheared,puring_frantar2023sparsegpt}, and knowledge distillation \cite{knowledge_liang2023less,knowledge_sahu2023promptmix,knowledge_gu2024minillm},  aim to reduce the memory footprint of LLMs, thus lowering the computational demands. Early Exit Strategies allow a model to terminate the computation at intermediate layers if certain conditions are met, thereby accelerating inference and reducing computational overhead. Early exit has been explored across various machine learning domains, focusing on designing efficient early exit networks \cite{earlyscore_bae-etal-2023-fast,earlyexitmodel_chen2023eellm}, and refining exit rules to balance accuracy and computational efficiency\cite{earlyscore_Zhou2020,earlyscore_Li2021}.

\section{Preliminaries}
\noindent\textbf{LLM Architecture and Decoding Process.} The typical architecture of LLMs sequentially consists of $N$ transformer layers and a language head, denoted as $\textrm{Head}(\cdot)$, for decoding the next token. Given an input sequence of tokens $\{x_1, \dots,  x_{t-1} \}$, the standard decoding process can be formally described as follows:
\begin{equation}
    p(x_{t}|x_{<t}) = \textrm{softmax}(\textrm{Head}(h^{(N)}_{t-1} ))_{x_{t}}
\end{equation}
where $h^{(N)}_{t-1}$ denotes the final hidden state output by the $N$-th (i.e., last) layer, while $p(x_{t}|x_{<t})$ represents the conditional distribution for sampling the next token.

\noindent\textbf{LLM-Based CTR Prediction. } We instruct LLMs to solve CTR prediction as a binary classification problem. Specifically, each task sample \(x_i\) is re-expressed into its natural language form $x^\text{text}_i$. Likewise, its corresponding binary label \(y_i \in \{0,1\}\) is mapped into a pair of binary response words $y^\text{text}_i \in$ \{\text{``yes''}, \text{``no''}\}. To obtain the LLM's tendencies between these two options, and consistent with prior work~\cite{WangLim2023,BaoEtAl2023,LinEtAl2024}, we consider only the binary response tokens (i.e., ``yes'' and ``no'') as candidates, excluding all other tokens in the vocabulary. This approach enables the extraction of the LLM's predictive tendencies between these two responses, as illustrated below:
\[
p(\textrm{yes}|x_{<t}) = \frac{\exp(s^{\textrm{yes}})}{\exp(s^{\textrm{yes}}) + \exp(s^{\textrm{no}})}.
\]
where \(s\) denotes the logit score for the given token, while \(p(\textrm{yes}|x_{<t})\) quantifies the LLM's preference for outputting the token ``yes''. Naturally, \(p(\textrm{yes}|x_{<t}) > 0.5\) is an intuitive and appropriate condition to finalize the outputted token.


\section{Retrieval Efficiency: GCN-Retriever}

To enable accurate learning and prediction, we construct textual sentences, denoted as \( X_{\text{text}} \) by integrating instructions, representative examples, and user input data. This process ensures the model aligns with the specific requirements and contexts of the task. For instance, in a product recommendation system, \( X_{\text{text}} \) is constructed as follows: 1) \textbf{Instruction:} Provide clear directives, such as "Predict whether the user will click on the given item." 2) \textbf{Examples:} include contextual examples, e.g., "User A rated Book X with 5 stars."  3) \textbf{Input:}  Incorporate real-time user interactions and queries. Complete prompt templates for each dataset are detailed in Appendix \ref{sec:prompts}.

To address the challenge of time efficiency in recommendation systems, we propose GCN-Retriever, a streamlined approach leveraging GCNs. This model constructs a bipartite graph where nodes represent users and items, and edges denote interactions between them.Through multi-layer message passing, GCN-Retriever refines user embeddings by incorporating neighborhood information. Specifically,
the embedding for a user \( u \)  at the next layer  \(k\) +1  is updated by aggregating features from connected neighbors: 


\[
e^{(k+1)}_u = \text{AGG}\left(e^{(k)}_u, \{e^{(k)}_i : i \in N_u\}\right)
\]
where \( e_u \) is the embedding of user \( u \), \( N_u \) denotes the neighbors of \( u \), and \text{AGG} represents the aggregation function which combines features of a node with those of its neighbors. 

To effectively capture the multidimensional signal of users, our GCN-Retriever model employs a strategy of averaging the embeddings obtained from different layers of GCNs. This approach provides a comprehensive representation that integrates diverse aspects of user behavior and attributes captured at the various levels of graph structure. The average of user embeddings is described by the following equation:
\[
\overline{e_u} = \frac{1}{K} \sum_{k=1}^K e_u^{(k)}.
\]
where \( \overline{e_u} \) represents the final averaged embedding for user \( u \), \( K \) is the number of layers from which embeddings are extracted and averaged, and \( e_u^{(k)} \) is the embedding of user \( u \) at layer \( k \). Finally, cosine similarity is used to measure the similarity between users based on their averaged embeddings. The complete algorithmic workflow, including embedding aggregation strategies and similar user retrieval process, is detailed in Algorithm 1 (see Appendix \ref{sec:algorithms}).
\section{Inference Acceleration: Dynamic Predictive Exiting}
In this section, we introduce \textbf{Dynamic Predictive Exiting} as a solution to Inference Slowdown. Motivated by \cite{earlyscore_Xin2021}, this mechanism leverages additional language heads to enable flexible inference termination while maintaining prediction quality. Specifically, during the forward pass through model layers, these language heads, attached to designated exit layers, decode the intermediate hidden states into next-token distributions. We design straightforward yet effective strategies to dynamically monitor the prediction confidence at different layers, using it as a real-time criterion to determine when to terminate inference and accept these intermediate distributions as final outputs.

This early exit methodology has proven effective in capturing the evolving dynamics of prediction preferences across different model layers in LLMs, functioning efficiently even without specialized training ~\citep{dola,earlyexit2,earlyexit3}. By applying the language head to the immature hidden states of the intermediate layers, we can calculate the probability of the next token solely conditioned on $h^{(j)}_{t-1}, j \in \{0, \dots, N-1 \}$, without finishing the entire inference process:
\begin{equation}
    p^{(j)}(x_{t}|x_{<t}) = \textrm{softmax}(\textrm{Head}(h^{(j)}_{t-1} ))_{x_{t}}
\end{equation}
Despite the advantages of this layer-wise predictive analysis in LLMs, it is widely recognized that intermediate hidden states typically present a significant information gap (i.e., distribution shift) compared to the final hidden states, leading to an unacceptable trade-off between efficency and response quality~\citep{prediction1,prediction2}. To address this issue, in the following section, we introduce additional language heads and propose a customized fine-tuning scheme for them. These fine-tuned language head can better ``understand'' the hidden states of earlier layers, thereby mitigating the information gap.

\subsection{Workflow of Dynamic Predictive Exiting}
\begin{figure*}[t]
  \includegraphics[width=1.0\linewidth]{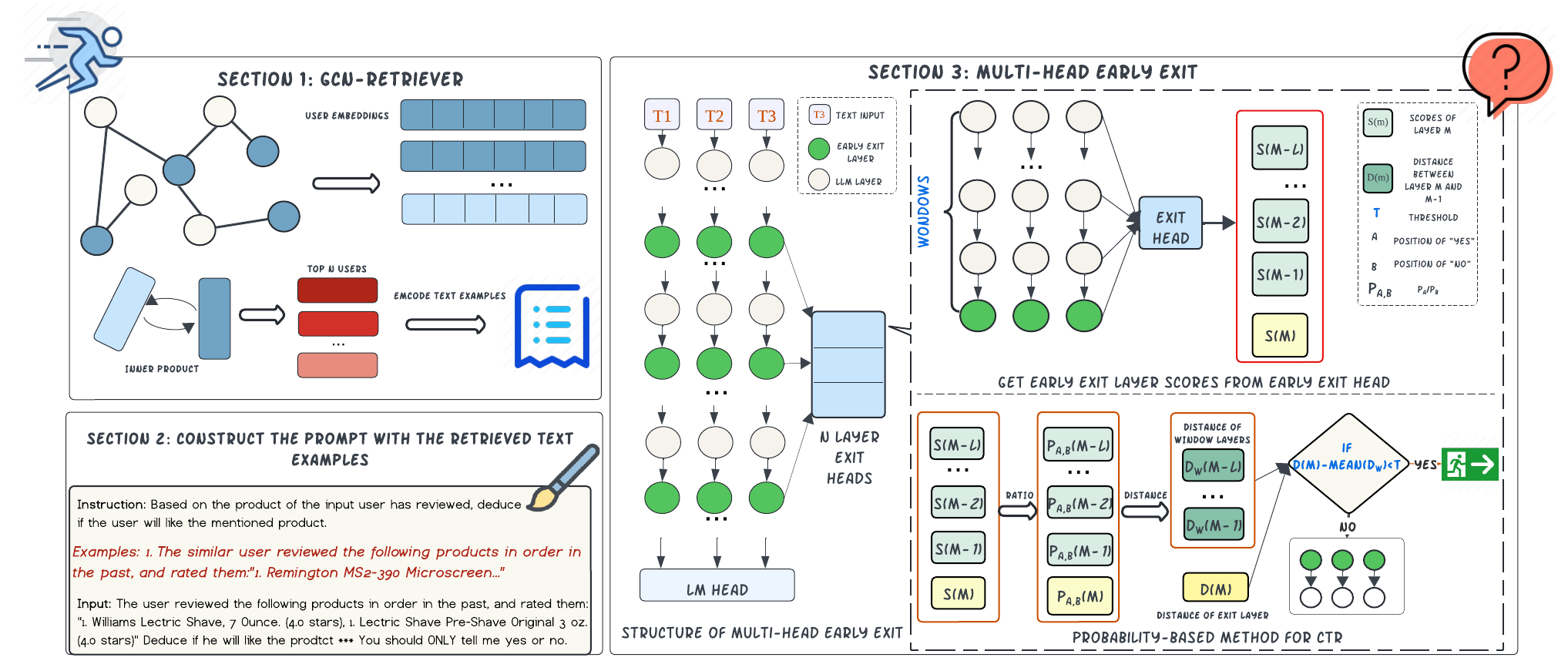}
  \caption{The process of OptiRAG-Rec: Section 1 GCN-Retriever: Constructs a GCN to generate user embeddings for identifying similar users; Section 2 Construct the Prompt with the Retrieved information: Forms LLM input by combining interaction data of the query user with similar users' data; Section 3 Multi-head Early Exit: Implements an early exit strategy in the LLM by designating potential exit layers and setting probability-based termination criteria.}
  \label{fig:framework_overview}
\end{figure*}


After obtaining the set of fine-tuned language head $\{\text{Head}_{\ell} | \ell \in \mathcal{L}\}$ at specified exit layers $\ell$, LLMs can decode intermediate hidden states in real-time to generate predictive distributions in advance. At this point, our proposed \textbf{Dynamic Predictive Exiting} mechanism can be applied to the target model for inference acceleration. The detailed workflow of this mechanism involves two steps as follows: 

\noindent \textbf{1. Dynamic real-time decoding.} 
When the forward computation of LLMs reaches the $\ell$-th  decoding layer,  we decode the hidden states $h_\ell$ in real-time to obtain the immature predictive distribution $P_\ell$, which reflects the prediction tendencies of the LLMs at the current layer.

\noindent \textbf{2. Predictive exiting strategies.} 
Each time we obtain the immature predictive distribution $P_\ell$ at an exiting layer $\ell$, a criterion is needed to decide whether to accept it as the final prediction and terminate early. To this end, we propose the strategy for LLM early exit for CTR. This  strategy are inspired by an interesting finding~\citep{dola} that LLMs progressively refine their hidden states across decoding layers. For some simple prediction steps, the hidden states at intermediate layers have already encoded sufficient information to predict the next token and remain relatively stable throughout the rest of the inference process. The goal of these strategies is to avoid unnecessary computations on such steps, thereby reducing the time spent on ``over-thinking''. Below, we detail the Probability-Based Method for CTR.



\textbf{Probability-Based Method for CTR.} 
We track the stability of the model's binary predictions (“yes” vs. “no”) across layers to determine early-exit readiness. Let \( k \in \{\text{yes}, \text{no}\} \) be the binary decision token and \( P_\ell(k) \) the predicted probability of token \( k \) at layer \( \ell \). The discrepancy in predictions between layer \( \ell \) and a previous layer \( \ell' \) is quantified by:
\[
D(P_\ell , P_{\ell'}) = \left| P_\ell(k) - P_{\ell'}(k) \right|,
\]

where \( D_{\ell\ell'} \in \mathbb{R} \) measures the absolute difference in predicted probabilities. In particular, the distance between consecutive layers \( \ell \) and \( \ell-1 \) is represented as \( D_\ell \). A smaller \( D_\ell \) indicates greater consistency in predictions across layers. To assess stability across multiple layers, we define the mean discrepancy over a window of \( m \) layers as:
\[
\overline{D}_\ell^m = \frac{1}{m} \sum_{i=\ell}^{\ell+m-1} D_i,
\]

If the absolute difference between the current discrepancy \( D_\ell \) and the average discrepancy over the window is below a predefined threshold \( \tau \), i.e.,

\[
\left| D_\ell - \overline{D}_{\ell-1}^m \right| < \tau,
\]

we consider the output at layer \( \ell \) to be stable and allow early termination of the forward pass. This logic is implemented in the multi-head early exit mechanism, as detailed in Algorithm~\ref{alg:early_exit} (Appendix~\ref{sec:algorithms}).


\subsection{Optimization of Multi Heads}
In prior to fine-tuning language models, we pre-define exit layers $\mathcal{L} \subset \{1, \dots, N-1 \}$ and mount an additional language head $\textrm{Head}_\ell$ at each exit layer $\ell \in \mathcal{L}$, initialized using the LLM's existing head. These heads serve two key purposes: decoding intermediate hidden states and approximating subsequent layer inferences, enabling high-quality predictive distributions from exited states.

Our fine-tuning process begins with comprehensive instruction-tuning of the entire vanilla model to align it with the target application. After fine-tuning, we freeze the model's parameters and integrate multi-early-exit heads (i.e., additional language heads) at the designated layers. Our empirical results show that fine-tuning only these heads improves training stability and leads to a better convergence rate. Specifically, for a given textual input $x$ and its ground truth token $y$, the layer-specific training loss for layer $\ell$ is defined as:
\[
L_{\mathrm{exit}}^{(\ell)} = -\log p^{(\ell)}(y|x),
\]
where \(p^{(\ell)}(y|x)\) is the probability that $\textrm{Head}_\ell$ assigns to the correct ground truth token \(y\).

In training multi-head architectures, a depth-varying learning rate strategy is beneficial. Shallower layers, capturing generic features, are assigned higher learning rates for aggressive updates, while deeper layers, adapting to specific features, benefit from finer updates. The learning rate for each head \( n \), located at depth \( d_n \) in the architecture, is defined as follows:

\[
\lambda_n = \lambda_0 \cdot e^{-\beta \cdot d_\ell},
\]

where \(\lambda_0\) is the base learning rate, \(\beta\)  controls the decay rate with depth, and \(d_\ell\) represents the head's depth, with shallower heads having a smaller \(d_\ell\).

\begin{table}[h]
\centering
\caption{Dataset Statistics}
\label{tab:dataset_stats}
\small
\begin{tabular}{@{}lccc@{}}
\toprule
Dataset & Users & Items & Samples \\
\midrule
Beauty & 324,037 & 32,892 & 6,525 \\
BookCrossing & 278,858 & 271,375 & 17,714 \\
Video Games & 37,890 & 17,381 & 221,465 \\
Movies and TV & 297,529 & 203,766 & 3,410,019 \\
Yelp & 228,195 & 146,927 & 2,956,589 \\
\bottomrule
\end{tabular}
\end{table}
 

  
  


\section{Experiments}
\textbf{Dataset.} We conduct experiments on three real-world datasets: BookCrossing\cite{bookcrossing}, Amazon Beauty, video games, Movies and TV\cite{amazonreview} and Yelp. We present the processed dataset statistics in Figure \ref{tab:dataset_stats}. 
\textbf{BookCrossing:} The BookCrossing dataset comprises user ratings and detailed textual descriptions of books. \textbf{Amazon dataset:} The Amazon dataset comprises user purchase actions and rating information sourced from the Amazon platform. For our experiments, we selected two domains with a substantial number of overlapping users: \textbf{Beauty} 2018, \textbf{Video Games} 2018 and \textbf{Movies and TV} 2018. The Yelp dataset includes user-generated ratings, reviews, and check-ins for businesses such as restaurants, which are collected from Yelp.com. To prepare the dataset for the recommender system experiments, we initially processed the original data by organizing past interactions chronologically for each user. We then filtered out samples that had fewer than three past interactions to ensure sufficient data quality and reliability in the training set. For the construction of recommendation tuning samples, we retained up to 15 interactions that occurred prior to the target item. We further binarize the ratings according to a threshold of 3. The refined dataset was subsequently divided into training, validation, and testing sets, maintaining a ratio of 8:1:1.


\begin{table*}[h]
\centering
\caption{Performance Comparison of Sequential Recommendation Models: Conventional Baselines, LLM Baseline, and Our Enhancements with GCN-Retriever and GCN-Early Exit.}
\label{tab:results}
\small 
\setlength{\tabcolsep}{3pt} 
\begin{tabular}{@{}ll|cc|cc|cc|cc|cc@{}}
\toprule
\multicolumn{2}{c}{\multirow{2}{*}{Model}}  & \multicolumn{2}{c}{BookCrossing} & \multicolumn{2}{c}{Beauty} & \multicolumn{2}{c}{Video Games} & \multicolumn{2}{c}{Movies and TV} & \multicolumn{2}{c}{Yelp}\\ 
 &  & AUC & Log Loss & AUC & Log loss & AUC & Log Loss & AUC & Log Loss & AUC & Log Loss \\ \midrule
\multirow{11}{*}{Full-shot} 
&  \multicolumn{1}{l|}{DeepFM}            & 71.15 & 0.6897 & 66.67 & 0.539 & 57.68 & 0.5916 & 80.33 & 0.3884 & 83.15 & 0.4279 \\
& DIN              & 71.17 & 0.6525 & 69.93 & 0.5727 & 62.96 & 0.7477 & 74.88 & 0.5210 & 71.59 & 0.7481 \\
& GRU4Rec          & 60.75 & 2.1302 & 64.82 & 0.6195 & 58.26 & 0.9163 & 70.64 & 0.5830 & 63.15 & 0.9439 \\
& AutoInt          & 58.00 & 0.6859 & 70.53 & 0.3899 & 63.59 & 0.8760 & 79.98 & 0.3902 & 83.02 & 0.4285 \\
& DCNv2            & 58.52 & 0.6758 & 69.37 & 0.5254 & 67.89 & 0.5502 & 79.53 & 0.3968 & 83.21 & 0.4264 \\
& WuKong           & 58.06 & 0.6798 & 65.64 & 0.3327 & 66.22 & 0.5926 & 77.36 & 0.4274 & 83.23 & 0.4261 \\  
& GDCN           & 58.03 & 0.6750 & 68.51 & 0.3587 & 66.27 & 0.7024 & 80.19 & 0.3922 & 83.08 & 0.4278 \\  
& EulerNet   & 57.86 & 0.7103 & 67.78 & 0.3212 &64.14 & 0.8886 & 79.19 & 0.3931 & 82.97 & 0.4273 \\ 
\midrule
\multirow{4}{*}{Few-shot} 
& TALLRec          & 70.74 & 0.6306 & 90.37 & 0.2459 & 75.41 & 0.4754 & 71.14 & 0.4514 & 77.38 & 0.4792 \\
& LLM-retriever    & 70.86 &  0.6907 & 89.65 & 0.2394 & 75.76 & 0.4805 & 73.89 & 0.7951 & 80.32 & 0.4577 \\
& \textbf{GCN-retriever} & 72.83 & 0.6158 & 94.72 & 0.2216 & 78.03 & 0.4850 & 90.34 &  0.4081 & 81.50 & 0.4692 \\
& \textbf{OptiRAG-Rec} & \textbf{82.11} &  \textbf{0.5269} &  \textbf{96.37} &  \textbf{0.2053} &  \textbf{97.86} &  \textbf{0.1911} &  \textbf{98.46} & \textbf{0.3010} &  \textbf{95.28} &  \textbf{0.2460} \\
\bottomrule
\end{tabular}
\end{table*}

\begin{table*}[h]
\centering
\caption{AUC Scores and Log Loss by Dataset and Retrieval Layer. }
\label{tab:gcn_layer}
\small
\begin{tabular}{@{}c|cc|cc|cc|cc|cc@{}}
\toprule
Retriever & \multicolumn{2}{c}{BookCrossing} & \multicolumn{2}{c}{Beauty} & \multicolumn{2}{c}{Video Games} & \multicolumn{2}{c}{Movies and TV} & \multicolumn{2}{c}{Yelp} \\
 & AUC  & Log Loss & AUC  & Log Loss & AUC  & LogLoss & AUC  & Log Loss & AUC  & Log Loss \\ \midrule
Average Layer  & 72.83  & \textbf{0.6158} & \textbf{94.72} & \textbf{0.2216} & \textbf{78.03} &  \textbf{0.4850} & 90.34 & \textbf{0.3850}  & \textbf{81.50} & 0.4692 \\
Last Layer     & 69.45  & 0.7151 & 93.55 &0.3774 & 70.88 & 0.5461 & \textbf{91.78} & 0.4062 & 80.95 & \textbf{0.4654} \\
Weighted Layer & \textbf{73.05} & 0.6293 & 93.64 & 0.3012 & 72.48 &  0.6573 & 89.86 &  0.4442 & 81.38  & 0.4935 \\
\bottomrule
\end{tabular}
\end{table*}

\begin{table*}[h]
\centering
\caption{Ablation Study: Component Analysis.}
\label{tab:Ablation}
\small
\begin{tabular}{@{}l|cc|cc|cc|cc|cc@{}}
\toprule
\multirow{2}{*}{Model Variant} & \multicolumn{2}{c|}{BookCrossing} & \multicolumn{2}{c|}{Beauty} & \multicolumn{2}{c|}{Video Games} & \multicolumn{2}{c|}{Movies and TV} & \multicolumn{2}{c}{Yelp} \\
 & AUC & RPS & AUC & RPS & AUC & RPS & AUC & RPS & AUC & RPS \\ 
\midrule
\multicolumn{11}{l}{\textit{Baseline}} \\
No Retriever, No Early Exit & 70.74 & \textbf{15.765} & 90.37 & \underline{15.098} & 75.41 & \textbf{8.879} & 71.14 & \textbf{7.585} & 77.38 & \textbf{11.486} \\
\midrule
\multicolumn{11}{l}{\textit{Single Component}} \\
Retriever Only & 78.03 & 3.825 & \underline{94.72} & 4.781 & 78.03 & 3.825 & \underline{90.34} & 3.674 & 80.34 & 3.821 \\
Early Exit Only & 75.98 & \underline{11.313} & 92.56 & \textbf{20.318} & \underline{96.37} & \underline{7.917} & \underline{97.40} & \underline{7.034} & 79.00 & \underline{8.216} \\
\midrule
\multicolumn{11}{l}{\textit{Combined Components}} \\
Single Head Early Exit & \underline{81.14} & 7.191 & 94.50 & 4.680 & 96.13 & 4.814 & \textbf{98.46} & 4.977 & \textbf{95.76} & 4.065 \\
Full Model (Both) & \textbf{82.11} & 5.505 & \textbf{96.37} & 4.960 & \textbf{97.86} & 4.932 & \textbf{98.46} & 5.080 & \underline{95.28} & 3.838 \\
\bottomrule
\multicolumn{11}{l}{\small Note: Four samples were retrieved for each dataset in the retriever. Best values are in \textbf{bold}, second-best are \underline{underlined}.} \\
\end{tabular}
\end{table*}

\begin{table*}[h]
\centering
\small
\caption{Performance comparison between GCN-retriever and FM-retriever across datasets.}
\begin{tabular}{l|cc|cc|cc|cc|cc}
\toprule
Model & \multicolumn{2}{c|}{BookCrossing} & \multicolumn{2}{c|}{Beauty} & \multicolumn{2}{c|}{Video Games} & \multicolumn{2}{c|}{Movies and TV} & \multicolumn{2}{c}{Yelp} \\
 & AUC & Log Loss & AUC & Log Loss & AUC & Log Loss & AUC & Log Loss & AUC & Log Loss \\
\midrule
GCN-retriever & \textbf{72.83} & \textbf{0.6158} & \textbf{94.72} & \textbf{0.2216} & \textbf{78.03} & \textbf{0.4850} & \textbf{90.34} & \textbf{0.4081} & \textbf{81.50} & \textbf{0.4692} \\
FM-retriever & 69.33 & 0.6409 & 88.03 & 0.6263 & 67.82 & 0.7443 & 73.17 & 0.6260 & 74.01 & 0.5484 \\
\bottomrule
\end{tabular}
\label{tab:gcn_vs_fm}
\end{table*}

\noindent\textbf{Baseline Methods.} Traditional CTR models are generally categorized into two types: feature interaction models and user behavior models. For our study, we selected DeepFM \cite{deepfm}, AutoInt\cite{autoint}, DCNv2\cite{wang2020dcn}, WuKong\cite{zhang2024wukong}, GDCN\cite{wang2023deeper} and EulerNet\cite{EulerNet} as representative feature interaction models. For user behavior models, we chose GRU4Rec \cite{gru4rec} and DIN \cite{din}. Additionally, we evaluated TALLRec \cite{tallrec} as a representative LLM-based CTR model. To further benchmark the efficiency-accuracy trade-off in LLM-based recommendation, we included two lightweight and compressed LLMs as additional baselines: \textbf{SparseGPT} \cite{sparsegpt}, a sparsity-aware pruned version of GPT for efficient inference, and \textbf{TinyLLaMA} \cite{tinyllm}, a distilled and highly compact LLaMA variant designed for low-resource deployment. These models allow us to evaluate performance under extreme efficiency constraints.

We also implemented an \textbf{FM-based retriever}, a classical factorization machine model \cite{fm} that scores user-item similarity based on learned embeddings. This provides a point of comparison to our GCN-retriever and highlights the benefits of graph-structured retrieval over traditional embedding similarity methods.

\noindent\textbf{Implementation Details.} For our experiments, we used Vicuna-7B, a model released by FastChat, as the base large language model. We employed few-shot training methods (randomly select less than 10\% of the training data) to fine-tune the model for CTR task, while using the entire training dataset for training the traditional models. In OptiRAG-Rec, the number of examples was set to four. To reduce computational overhead in calculating exit scores, we selected layers (5,10,15, 20, 25, 30) as the early exit layers. For the multi-head early exit mechanism, we used a default window size of 3 and a threshold of 0.01. These values were selected for consistency across datasets in the main comparison. Additional ablation experiments with varying window sizes (3, 4, 5) and thresholds (0.005, 0.01, 0.05) are provided in Appendix~\ref{sec:hyperparameter_analysis}. For the GCN-retriever, we configured the model with 3 layers, an embedding dimension of 64, and trained it until the evaluation loss converged, ensuring optimal performance for recommendation. All methods used identical input features and preprocessing. Hyperparameter tuning followed consistent procedures across all baselines to ensure fair comparison. The superior LLM performance despite using only 10\% training data validates the effectiveness of our approach under practical computational constraints.

\noindent\textbf{Measurement.} Each configuration's performance was assessed using the area under the curve (AUC) and log loss for accuracy. The retrieval times, indicating computational demand, were normalized to the baseline (1x) set by the LLM retrievor. Inference speed was measured in terms of requests per second (RPS) per NVIDIA A100 GPU.

\begin{figure}[h]
  \centering
  \begin{minipage}{0.99\linewidth} 
    \centering
\includegraphics[width=0.80\linewidth]{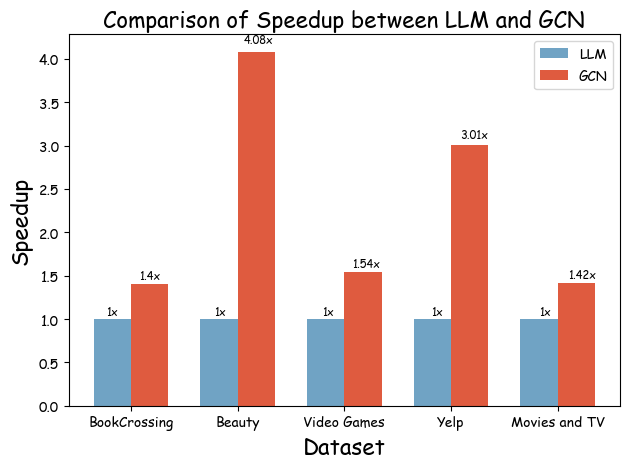} 
    \caption{Speedup Comparison between LLM and GCN by Dataset.}
    \label{fig:retrieve time}
  \end{minipage}
\end{figure}

\begin{figure*}[h]
  \centering
  \begin{minipage}{0.99\linewidth} 
    \centering
    \includegraphics[width=0.99\linewidth]{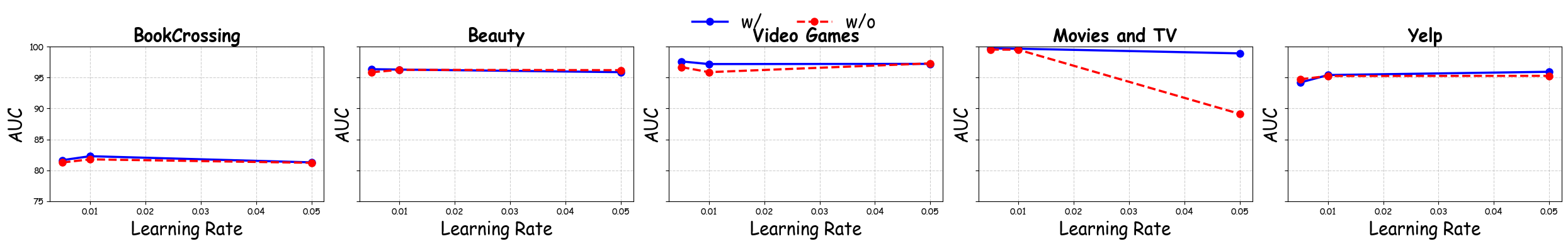} 
    \caption{Comparison of model performance with (w/) and without (w/o) adaptive learning rate. }
    \label{fig: lr}
  \end{minipage}
\end{figure*}

\textbf{Q1: Does OptiRAG-Rec Outperform traditional recommendation Models?} Table \ref{tab:results} highlights key findings: OptiRAG-Rec significantly outperforms traditional methods, with AUC improvements of 16.14 for BookCrossing, 11.80 for Yelp, and 19.59 for Movies and TV(averaged across traditional baselines). This underscores the efficacy of our integrated framework combining GCN-based retrieval with LLM processing. Compared to traditional models, LLM-based methods excel in CTR prediction, especially for text-rich datasets like BookCrossing and Amazon reviews, which include item details such as titles, years, and prices. Unlike existing LLM-based approaches like TallRec, our method incorporates similar users' interactions, offering a more comprehensive view of user preferences and outperforming user-only LLM frameworks in CTR tasks.

\textbf{Q2: What Are the Advantages of Multi-Head Early Exit Over Baselines and Efficiency Alternatives?}
Building on the overall framework success, we analyze the specific contribution of our novel multi-head early exit mechanism, a key component of OptiRAG-Rec's efficiency optimization.
\noindent
We evaluate the effectiveness of this approach by comparing it against (1) a baseline model without early exit, (2) a single-head early exit variant, and (3) other efficiency methods such as SparseGPT and TinyLLaMA. As shown in Table~\ref{tab:Ablation}, the multi-head early exit model achieves consistently strong results across datasets. For example, on the Beauty dataset, it achieves an AUC of \textbf{96.37} with an RPS of 4.960, outperforming both the single-head variant (AUC 94.50, RPS 4.680) and the no-exit baseline.

Comparison with other efficiency methods (detailed results in Appendix~\ref{sec:efficiency_analysis}) reveals that our architectural approach outperforms traditional compression techniques. For instance, our method achieves 96.37 AUC vs 92.78 for SparseGPT on Beauty, while maintaining competitive throughput. Unlike quantization or pruning methods that reduce model capacity, our multi-head early exit preserves full precision while achieving efficiency through intelligent termination decisions.



\noindent
We attribute the effectiveness of our design to two core factors:
\begin{itemize}
    \item \textit{Layer-wise specialization:} Each exit head is independently fine-tuned at different layers, enabling early predictions for simpler inputs while allowing deeper computation for more complex cases.
    \item \textit{Robust confidence estimation:} A smoothed token-level scoring mechanism assesses prediction stability over time, supporting reliable and adaptive exit decisions.
\end{itemize}
\noindent
The multi-head architecture provides superior reliability by aggregating confidence signals from multiple specialized perspectives, reducing the variance in early exit decisions that can occur with single-head approaches. These results validate that a multi-head architecture, when paired with a tailored exit criterion, offers a principled and effective approach to balancing computational efficiency and predictive performance in LLM-based recommendation systems.

\textbf{Q3: Why is GCN-Based Retrieval Superior to LLM and Direct Similarity Methods?}
Our analysis (Figure \ref{fig:retrieve time} and Table \ref{tab:results}) shows that GCN-retrievers achieve significantly faster retrieval speeds and higher accuracy than LLM-based retrievers. For instance, on the BookCrossing dataset, the GCN-retriever achieves an AUC of 82.11, outperforming the LLM retriever's 72.83. This performance gap reflects fundamental architectural differences between the two approaches:

\begin{itemize}
    \item \textit{Information Coverage:} LLM-based retrievers are constrained by limited context windows, incorporating only the 15 most recent interactions due to computational and memory limitations in our analysis. This truncation hinders the model’s ability to model long-term user preferences. In contrast, GCN retrievers operate over the entire user-item interaction graph, leveraging multi-hop connectivity and graph structure to identify latent community-level patterns and indirect user-item relationships.
    
    \item \textit{Computational Efficiency:} GCN retrievers use lightweight graph operations with linear complexity, while LLM retrievers require expensive transformer computations that scale quadratically with input length, adding overhead to similarity calculations.
\end{itemize}




To further isolate the contribution of graph topology, we compared our GCN-retriever with an FM-based retriever, which relies on direct user-item similarity scores without structural modeling. As shown in Table~\ref{tab:gcn_vs_fm}, GCN-retriever consistently outperforms FM across all datasets, with AUC gains ranging from 2.68 to 17.17 points. These results confirm that the observed improvements are not merely due to representation learning, but stem from the ability of graph-based methods to capture complex, indirect user-item relationships via the interaction network.

\textbf{Q4: How Do Different GCN Layer Aggregation Strategies Compare?}
Our analysis (Table \ref{tab:gcn_layer}) reveals that averaging embeddings from multiple GCN layers produces more robust representations compared to using either the final layer  or weighted embeddings. This suggests that embeddings averaged across multiple layers provide a richer, more generalized representation that captures a broader spectrum of user-item interaction patterns.

\textbf{Q5: What is the Impact of Combining Retrieval with Early Exit?}
We analyze the interaction effects between retrieval mechanisms and early exit strategies by examining four system configurations in Table~\ref{tab:Ablation}: baseline (neither), retrieval only, early exit only, and the combined approach.

The results demonstrate clear synergistic benefits when combining both components. The combination substantially improves computational throughput across all datasets, with Video Games showing RPS increases from 3.825 (retrieval only) to 4.932 (combined), representing a 28.9\% improvement. Similarly, Beauty demonstrates RPS improvement from 4.781 to 4.960, indicating that early exit effectively mitigates the computational overhead introduced by retrieval processing.

Notably, these efficiency gains do not come at the expense of accuracy. In fact, combining retrieval and early exit often enhances both metrics simultaneously. On Beauty, AUC increases from 94.72 (retrieval only) to 96.37 (combined), and on Movies and TV from 90.34 to 98.46. These improvements suggest that retrieval enhances input quality, enabling more confident and earlier predictions by the exit mechanism.

Importantly, the benefits of the combined system exceed the additive effects of the individual components. Retrieval improves representation quality, while early exit reduces inference cost. Their integration addresses complementary bottlenecks, resulting in a more efficient and accurate system than either component achieves in isolation.

\textbf{Q6: How Do Key Hyperparameters Affect Performance?} The Figure \ref{fig: lr} show that the adaptive learning rate (blue line) does not significantly outperform the non-adaptive learning rate (red line), as the AUC scores remain closely aligned across all datasets. This suggests that the adaptive learning rate may not substantially enhance performance for the tested data and model configurations. However, the consistent performance indicates the model's robustness to learning rate variations, which could be advantageous for maintaining stability across different operational conditions.

\section{Conclusion}
As LLMs continue to advance, their potential to transform recommendation systems is becoming increasingly evident \cite{tallrec}. In this work, we introduced a comprehensive framework that integrates advanced retrieval mechanisms and early exit strategies to enhance both the efficiency and accuracy of LLM-based recommendations. By incorporating Graph GCNs as efficient retrieval mechanisms and implementing a multi-head early exit architecture, our framework significantly reduces computation time while maintaining or even improving system accuracy. This holistic approach not only accelerates the responsiveness of LLMs but also preserves their decision-making quality, making it highly suitable for real-time applications in commercial systems. Our results demonstrate the effectiveness of this framework in balancing performance and efficiency, setting a new standard for deploying LLMs in recommendation tasks.

\section{Limitation}
Our framework has several limitations requiring future consideration. The approach requires substantial textual metadata, limiting applicability to sparse-text domains. The GCN-retriever struggles with cold-start scenarios lacking sufficient interaction history. Despite efficiency improvements, initial training requires significant computational resources. Domain transferability beyond e-commerce and reviews needs additional validation. The framework lacks real-time graph update mechanisms for continuously evolving interactions.


\clearpage
\newpage
\bibliographystyle{assets/plainnat}
\bibliography{paper}

\clearpage
\newpage
\beginappendix

\section{Appendix}
\subsection{Prompt Templates and Examples}
\label{sec:prompts}

Our framework uses dataset-specific prompt templates that integrate similar users' interactions with target user histories. Tables~\ref{tab:amazon_prompt}, \ref{tab:bookcrossing_prompt}, and \ref{tab:yelp_prompt} show representative examples from each dataset type. Note that Amazon datasets (Beauty, Video Games, Movies and TV) share the same prompt structure, differing only in product-specific details, so we present only the Beauty example as representative of all Amazon domains.

\begin{table}[h]
\setlength{\tabcolsep}{3.9pt}
\renewcommand{\arraystretch}{1.05}
\centering
\small
\begin{footnotesize}
\begin{tabularx}{0.48\textwidth}{X} 
\toprule
\textbf{Instruction:} Based on the product of the input user has reviewed, deduce if the user will like the mentioned product. Note that more stars the user rated the product, the user liked the product more. You should ONLY tell me yes or no.
\newline
\newline
\textbf{Similar Users' Examples:}
\newline
\texttt{1.  The similar user reviewed the following products and rated them:  [`Mederma Scar Gel, 20 Grams. (5.0 stars)', `Avalon Organics Wrinkle Therapy CoQ10 Cleansing Milk, 8.50 oz. It's price is \$8.27 (5.0 stars)', ...]}
\newline
\texttt{2. The similar user reviewed the following products and rated them: [`Caswell-Massey - Newport Soap on a Rope. (5.0 stars)', `Deep Steep 3 Piece Gift Set, Grapefruit Bergamot. (5.0 stars)', ...]}
\newline
\texttt{3. The similar user reviewed the following products and rated them: [`Dinur Cosmetics Bio Clean Drying Lotion 0.67 oz. (5.0 stars)', `Avalon Organics Vitamin C Renewal Creme, 2 oz. (5.0 stars)', ...]}
\newline
\texttt{4. The similar user reviewed the following products and rated them: [`Plant Therapy Tea Tree Organic Essential Oil (5.0 stars)', `Avalon Organics Vitamin C Renewal Moisture Plus Lotion SPF 15 (1.0 star)', ...]}
\newline
\newline
\textbf{Target User Input:} The user reviewed the following products in order in the past, and rated them: [`Avalon Organics Wrinkle Therapy CoQ10 Cleansing Milk, 8.50 oz. (5.0 stars)', `Organic Fiji Raw Organic Coconut Oil, 13-Ounce Jars (5.0 stars)', ...]
\newline
\newline
\textbf{Question:} Deduce if he will like the product ***(3 Pack) KLEANCOLOR Retractable Waterproof Lip \& Eye Liner - Fuschia***. You should ONLY tell me yes or no.
\\
\bottomrule
\end{tabularx}
\end{footnotesize}
\caption{Amazon Beauty dataset prompt example.}
\label{tab:amazon_prompt}
\end{table}

\begin{table}[h]
\setlength{\tabcolsep}{3.9pt}
\renewcommand{\arraystretch}{1.05}
\centering
\small
\begin{footnotesize}
\begin{tabularx}{0.48\textwidth}{X} 
\toprule
\textbf{Instruction:} Deduce if the user will like the candidate book based on the user's past history. You can refer to the books rating history of other users. Note that more stars the user rated the book, the user liked the book more. You should ONLY tell me yes or no. 
\newline
\newline
\textbf{Similar Users' Reading History:}
\newline
\texttt{1. The user's location is USA. The user's age is unknown. The user read the following books and rated them: [`Face the Fire (Three Sisters Island Trilogy) (6 stars)', `The Sigma Protocol (9 stars)', ...]}
\newline
\texttt{2. The user's location is Netherlands. The user's age is 35-39. The user read the following books and rated them: [`The Home DIY Expert (7 stars)', `Firefly Summer (7 stars)', `The Girls' Guide to Hunting and Fishing (9 stars)', ...]}
\newline
\texttt{3. The user's location is USA. The user's age is unknown. The user read the following books and rated them: [`Daughter of Fortune (8 stars)', `Ship of Gold in the Deep Blue Sea (5 stars)', `Thin Air (5 stars)', ...]}
\newline
\texttt{4. The user's location is USA. The user's age is 25-29. The user read the following books and rated them:[`Chasing the Dime (9 stars)', `Border Bride (10 stars)', `Murder at the Library of Congress (8 stars)', ...]}
\newline
\newline
\textbf{Target User Input:} The user's location is USA. The user's age is 45-49. The user read the following books and rated them: [`Dress Codes: Of Three Girlhoods (8 stars)', `What Would Buddha Do? (5 stars)', `Are You Somebody? (8 stars)', ...]
\newline
\newline
\textbf{Question:} Based on the books the user has read, deduce if the user will like the book ***Big Fish***. You should ONLY tell me yes or no.
\\
\bottomrule
\end{tabularx}
\end{footnotesize}
\caption{BookCrossing dataset prompt example.}
\label{tab:bookcrossing_prompt}
\end{table}

\begin{table}[h]
\setlength{\tabcolsep}{3.9pt}
\renewcommand{\arraystretch}{1.05}
\centering
\small
\begin{footnotesize}
\begin{tabularx}{0.48\textwidth}{X} 
\toprule
\textbf{Instruction:} Based on the Yelp store the user has visited, determine if the user will like the mentioned store. Note: The more stars the user rated the store, the more they liked it.You should ONLY tell me yes or no.
\newline
\newline
\textbf{Similar Users' Store Visits:}
\newline
\texttt{1. The similar user joined Yelp in 2016-07-20. They have 0 fans and typically rate stores with 3.82 stars average. The user recently visited the following stores and rated them: [`Matt \& Marie's, Philadelphia PA (1.0 star)', `The Caketeria, Newtown PA (5.0 stars)', `Kensington Quarters, Philadelphia PA (5.0 stars)', ...]}
\newline
\texttt{2. The similar user joined in 2011-02-21. They have 27 fans and rate at 3.86 stars average. The user recently visited the following stores and rated them: [`Brick House Tavern + Tap, Tampa FL (3.0 stars)', `Brio Italian Grille, Tampa FL (5.0 stars)', `Kona Grill - Tampa, Tampa FL (5.0 stars)', ...]}
\newline
\texttt{3. The similar user joined in 2014-06-04. They have 15 fans and rate at 3.97 stars average. The user recently visited the following stores and rated them: [`Mini Doughnut Factory, Tampa FL (5.0 stars)', `Green Lemon, Tampa FL (5.0 stars)', `Catch Twenty Three, Tampa FL (5.0 stars)', ...]}
\newline
\texttt{4. The similar user joined in 2014-09-27. They have 21 fans and rate at 4.42 stars average. The user recently visited the following stores and rated them: [`The Red Lion Pub, Indian Rocks Beach FL (4.0 stars)', `Hai Street Kitchen \& Co, Philadelphia PA (5.0 stars)', `The Mütter Museum, Philadelphia PA (4.0 stars)', ...]}
\newline
\newline
\textbf{Target User Input:} The user has been on Yelp since 2016-07-20. They have 0 fans and rate stores at 3.82 stars average. The user recently visited the following stores and rated them:[`Cross Culture, Doylestown PA (3.0 stars)', 'Dan Dan, Philadelphia PA (5.0 stars)', `Giwa Korean Kitchen, Philadelphia PA (5.0 stars)', ...]
\newline
\newline
\textbf{Task:} Deduce if the user will like the store ***Jamba located in Willow Grove PA***. Remember: You should ONLY respond with yes or no.
\\
\bottomrule
\end{tabularx}
\end{footnotesize}
\caption{Yelp dataset prompt example.}
\label{tab:yelp_prompt}
\end{table}

\subsubsection{Prompt Engineering Considerations}

\textbf{Retrieval Integration:} We retrieve 4 similar users to provide relevant context while maintaining manageable prompt length.

\textbf{History Limitation:} Target user history is limited to 15 most recent interactions to stay within LLM context windows while preserving recent preference signals.

\textbf{Output Constraint:} We restrict LLM output to binary tokens (``yes''/``no'') and extract prediction probabilities from logit scores, enabling efficient CTR prediction without full text generation.

\begin{algorithm}[H]
\caption{GCN-Based User Retriever}
\label{alg:gcn_retriever}
\begin{algorithmic}[1]
\small
\REQUIRE $\mathcal{H}$ (user histories), METHOD $\in$ \{`MEAN', `FINAL', `WEIGHT'\}, TOP\_K
\STATE $\mathcal{G} \leftarrow$ \text{GCN}($\mathcal{H}$)
\STATE $\mathbf{H}_u^{(l)} \leftarrow \mathcal{G}.\text{get\_embeddings()}$ for all users $u \in U$ and layers $l = 1, \dots, L$

\STATE // Aggregate layer-wise embeddings
\IF{METHOD == `MEAN'}
    \STATE $\mathbf{H}_u \leftarrow \frac{1}{L} \sum_{l=1}^{L} \mathbf{h}_u^{(l)}$
\ELSIF{METHOD == `FINAL'}
    \STATE $\mathbf{H}_u \leftarrow \mathbf{h}_u^{(L)}$
\ELSE
    \STATE $\mathbf{H}_u \leftarrow \sum_{l=1}^{L} \alpha_l \mathbf{h}_u^{(l)}$ \COMMENT{$\alpha_l$ are layer weights}
\ENDIF

\STATE // Normalize user embeddings
\STATE $\mathbf{H}_u \leftarrow \frac{\mathbf{H}_u}{\|\mathbf{H}_u\|_2}$ for all $u \in U$

\STATE // Generate prompts using top-k similar users
\STATE prompts $\leftarrow$ [ ]
\FOR{each user $u \in U$}
    \STATE $\mathbf{s}_u \leftarrow \mathbf{H}_u \cdot \mathbf{H}_U^\top$ 
    \STATE $\mathbf{s}_u[u] \leftarrow -\infty$ \COMMENT{Exclude self-matching}
    \STATE similar\_users $\leftarrow \text{ArgSort}(\mathbf{s}_u)[:\text{TOP\_K}]$
    \STATE prompt $\leftarrow \text{Format}(u, \text{similar\_users}, \mathcal{H})$
    \STATE prompts.append(prompt)
\ENDFOR
\RETURN prompts
\end{algorithmic}
\end{algorithm}





    


\begin{algorithm}[H]
\caption{Multihead Early-Exit}
\label{alg:early_exit}
\small
\begin{algorithmic}[1]

\REQUIRE $\mathcal{L}$: Exit layers; $h^{(\ell)}$: hidden states; $\textrm{Head}_\ell$: classifier head at layer $\ell$;\\
\hspace{1.9em} $m$: window size $(\geq 3)$; $\tau$: threshold for exit

\ENSURE \texttt{True} if early exit condition is met, else \texttt{False}

\STATE Initialize multiheads: $\{\textrm{Head}_\ell \leftarrow \textrm{Copy}(\textrm{LM Head}) \mid \ell \in \mathcal{L} \}$
\FOR{each $\ell \in \mathcal{L}$}
    \STATE Select $m{+}1$ most recent hidden states: $\mathcal{H}_\ell \leftarrow \{h^{(\ell-m)}, ..., h^{(\ell)}\}$
    \STATE Initialize empty ratio list $\mathcal{R}$
    \FOR{each $h \in \mathcal{H}_\ell$}
        \STATE $\mathbf{p} \leftarrow \textrm{Softmax}(\textrm{Head}_\ell(h))$
        \STATE Compute ratio: $r \leftarrow \frac{p_{\text{pos}}}{p_{\text{neg}}}$
        \STATE Append: $\mathcal{R}.\text{append}(r)$
    \ENDFOR

    \STATE Compute deltas: $\delta_i = |\mathcal{R}[i] - \mathcal{R}[i+1]|$ for $i = 0$ to $m{-}1$
    \STATE $\overline{D}_\ell^m \leftarrow \frac{1}{m-1} \sum_{i=0}^{m-2} \delta_i$
    \STATE $D_\ell \leftarrow |\mathcal{R}[-1] - \mathcal{R}[-2]|$

    \IF{$|D_\ell - \overline{D}_\ell^m| < \tau$}
        \RETURN \texttt{True}
    \ENDIF
\ENDFOR

\RETURN \texttt{False}
\end{algorithmic}
\end{algorithm}

\subsection{Algorithmic Details}
\label{sec:algorithms}
Algorithm~\ref{alg:early_exit} and Algorithm~\ref{alg:gcn_retriever} provide detailed pseudocode for our core components.

\subsection{Comprehensive Efficiency Analysis}
\label{sec:efficiency_analysis}
To address concerns about efficiency trade-offs, we conducted comprehensive experiments comparing OptiRAG-Rec with prominent alternative efficiency techniques. Table~\ref{tab:efficiency_accuracy} presents detailed results across all datasets.

\subsubsection{Experimental Setup}

We evaluated two efficiency approaches:

\textbf{SparseGPT + GCN-Retriever:} We applied structured pruning via SparseGPT~\cite{sparsegpt} to our backbone LLM while maintaining the GCN-Retriever architecture. Specifically, we introduced a sparsity level of 0.5 for the projection layers (`q\_proj', `k\_proj', `v\_proj', `o\_proj') in each transformer layer, targeting the most computationally intensive components.

\textbf{TinyLlama + Fine-tuned GCN-Retriever:} We replaced the backbone LLM with TinyLlama~\cite{tinyllm} while retaining our GCN-Retriever component, representing the model compression approach to efficiency optimization.

\subsubsection{Key Findings and Analysis}



\begin{figure*}[h]
\centering
\small
\includegraphics[width=0.8\linewidth]{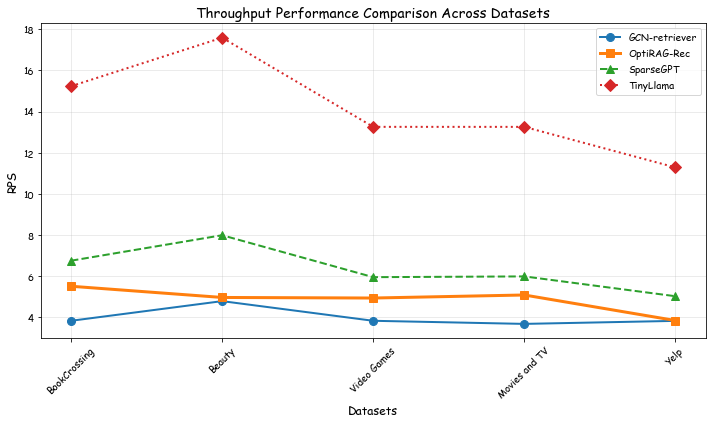}
\caption{Throughput Performance Comparison (RPS) Across Datasets}
\label{fig:throughput_comparison}
\end{figure*}

\textbf{Pruning Trade-offs:} As shown in Tables~\ref{tab:efficiency_accuracy} and~\ref{fig:throughput_comparison}, SparseGPT achieves higher throughput (up to 7.99 RPS vs 4.96 on Beauty) but suffers significant accuracy degradation. BookCrossing shows a 15.27-point AUC decrease (82.11 vs 66.84), demonstrating that static pruning cannot preserve the quality-efficiency balance of dynamic computation.

\textbf{Model Compression Limitations:} TinyLlama achieves the highest throughput (up to 17.60 RPS) but experiences catastrophic accuracy loss of 30-45\% compared to OptiRAG-Rec. This reveals fundamental limitations of capacity reduction for recommendation requiring sophisticated reasoning.

\textbf{Dynamic Efficiency Advantage:} OptiRAG-Rec simultaneously enhances accuracy (9.28-point average improvement over baseline) while maintaining competitive throughput (3.84-5.51 RPS). Unlike static methods with fixed trade-offs, our dynamic approach adapts computational allocation based on sample complexity.


These results validate our design philosophy: architectural innovations in dynamic computation allocation provide superior efficiency gains compared to traditional model compression approaches. OptiRAG-Rec offers the optimal balance for production recommendation systems where both accuracy and latency are critical constraints, while alternative methods force suboptimal trade-offs between these competing objectives.

\subsection{Hyperparameter Sensitivity Analysis}
\label{sec:hyperparameter_analysis}
\begin{figure*}
    \centering
    \caption{Hyperparameter Sensitivity Analysis across Different Configurations}
    \label{fig:Combined}
    \begin{subfigure}[t]{0.32\textwidth}
        \centering
        \includegraphics[width=\linewidth]{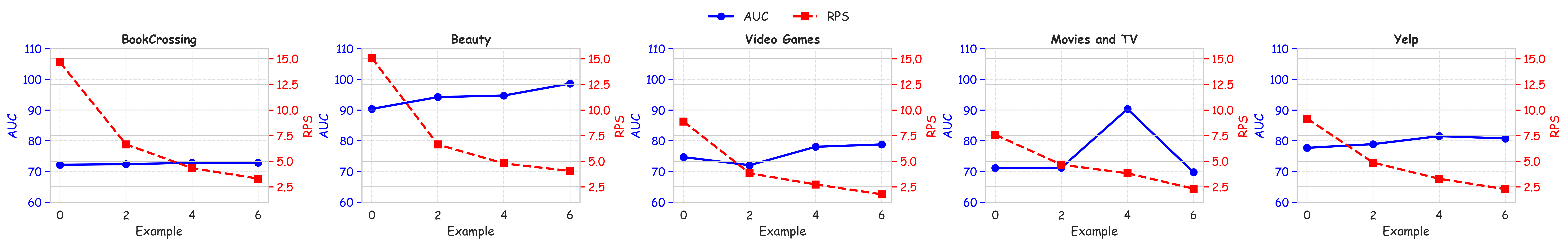}
        \caption{AUC and RPS by retrieval examples.}
        \label{fig:Example}
    \end{subfigure}
    \hfill
    \begin{subfigure}[t]{0.32\textwidth}
        \centering
        \includegraphics[width=\linewidth]{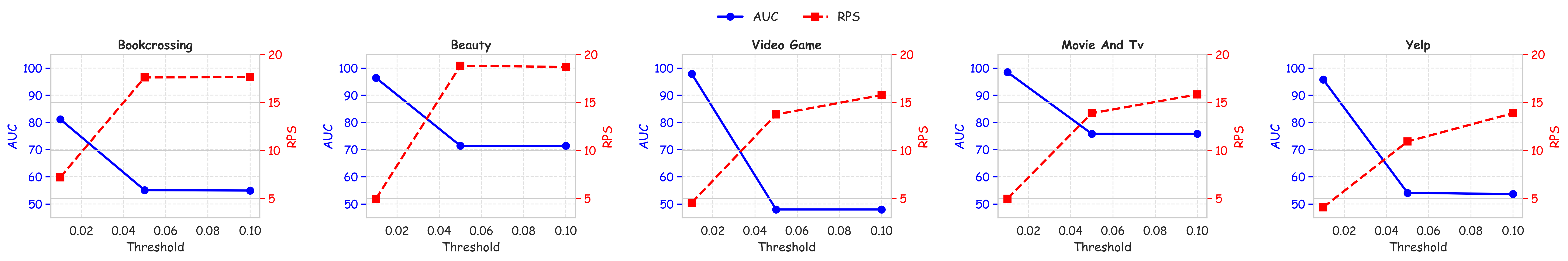}
        \caption{AUC and RPS by threshold.}
        \label{fig:Threshold}
    \end{subfigure}
    \hfill
    \begin{subfigure}[t]{0.32\textwidth}
        \centering
        \includegraphics[width=\linewidth]{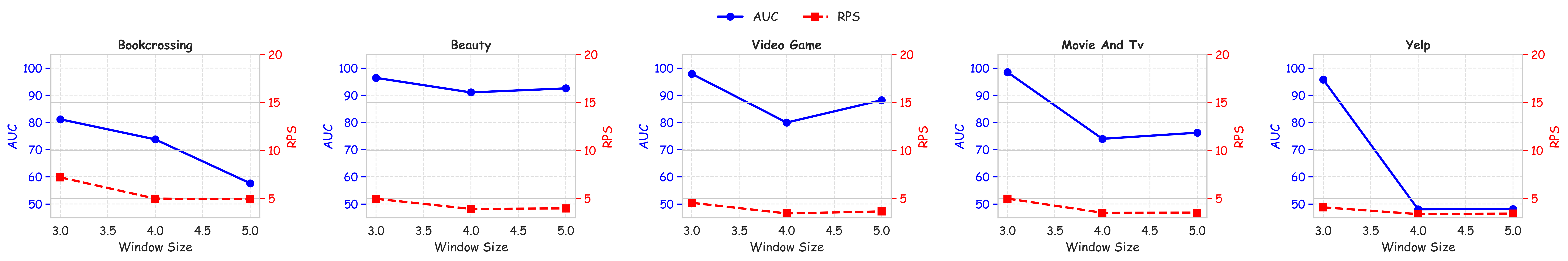}
        \caption{AUC and RPS by windows.}
        \label{fig:Windows}
    \end{subfigure}
\end{figure*}

Figure~\ref{fig:Combined} presents comprehensive analysis of key hyperparameters affecting system performance.

\textbf{Retrieval Examples Impact:} Figure~\ref{fig:Example} shows the trade-off between retrieval quantity and performance. AUC generally improves with more examples up to 4-5, after which diminishing returns occur. This indicates an optimal balance between context richness and computational overhead, with 4 examples providing the best accuracy-efficiency trade-off across datasets.

\textbf{Early Exit Configuration:} Figures~\ref{fig:Threshold} and~\ref{fig:Windows} show the effects of varying early exit thresholds and window sizes. Results show that using a threshold of 0.01 with a window size of 3 consistently achieves the best trade-off between accuracy and efficiency (e.g., 96.37 AUC on BookCrossing, 97.86 on Video Games). Higher thresholds (0.05, 0.1) lead to premature exits with poor accuracy, while larger windows reduce throughput without clear gains. We adopt threshold=0.01 and window=3 as our default setting.

\begin{table*}[h]
\centering
\caption{Efficiency Method Comparison: Accuracy and Quality Metrics}
\label{tab:efficiency_accuracy}
\small
\begin{tabular}{@{}l|cc|cc|cc|cc|cc@{}}
\toprule
\multirow{2}{*}{Method} & \multicolumn{2}{c|}{BookCrossing} & \multicolumn{2}{c|}{Beauty} & \multicolumn{2}{c|}{Video Games} & \multicolumn{2}{c|}{Movies and TV} & \multicolumn{2}{c}{Yelp} \\
 & AUC & Log Loss & AUC & Log Loss & AUC & Log Loss & AUC & Log Loss & AUC & Log Loss \\
\midrule
GCN-retriever & 72.83 & 0.6158 & 94.72 & 0.2216 & 78.03 & 0.4850 & 90.34 & 0.4081 & 81.50 & 0.4692 \\
\textbf{OptiRAG-Rec} & \textbf{82.11} & \textbf{0.5269} & \textbf{96.37} & \textbf{0.2053} & \textbf{97.86} & \textbf{0.1911} & \textbf{98.46} & \textbf{0.1911} & \textbf{95.28} & \textbf{0.2460} \\
SparseGPT & 66.84 & 0.6632 & 92.78 & 0.2631 & 69.59 & 0.6348 & 90.61 & 0.3854 & 77.48 & 0.3854 \\
TinyLlama & 52.81 & 0.8754 & 66.61 & 0.3699 & 54.71 & 0.5534 & 78.86 & 0.4458 & 50.70 & 0.4456 \\
\bottomrule
\end{tabular}
\end{table*}

\end{document}